\documentclass[aps,prl,10pt,twocolumn,tightenlines,amsmath,amssymb,final,letterpaper,superscriptaddress,notitlepage,nofootinbib,longbibliography]{revtex4-1}
\usepackage{graphicx}
\usepackage{subfigure}
\usepackage{epstopdf}
\usepackage{times}
\usepackage{bm}
\usepackage[dvipsnames]{xcolor}
\usepackage{tcolorbox}
\usepackage[hidelinks]{hyperref}

\graphicspath{{figs/}} 

\begin{document}
%Working title
\title{Self-reinforcing cascades: A spreading model for beliefs or products of varying intensity or quality}

\author{Laurent \surname{H\'ebert-Dufresne}}
\affiliation{Vermont Complex Systems Institute, University of Vermont, Burlington VT, USA}
\affiliation{Department of Computer Science, University of Vermont, Burlington VT, USA}
\affiliation{Santa Fe Institute, 1399 Hyde Park Road, Santa Fe, NM 87501, USA}

\author{Juniper \surname{Lovato}}
\affiliation{Vermont Complex Systems Institute, University of Vermont, Burlington VT, USA}
\affiliation{Department of Computer Science, University of Vermont, Burlington VT, USA}

\author{Giulio \surname{Burgio}}
\affiliation{Vermont Complex Systems Institute, University of Vermont, Burlington VT, USA}

\author{James P. \surname{Gleeson}}
\affiliation{MACSI, Department of Mathematics and Statistics, University of Limerick, Limerick, Ireland}

\author{S. Redner}
\affiliation{Santa Fe Institute, 1399 Hyde Park Road, Santa Fe, NM 87501, USA}
\affiliation{Vermont Complex Systems Institute, University of Vermont, Burlington VT, USA}

\author{P. L. Krapivsky}
\affiliation{Department of Physics, Boston University, Boston, MA 02215, USA}
\affiliation{Santa Fe Institute, 1399 Hyde Park Road, Santa Fe, NM 87501, USA}
\begin{abstract}
Models of how things spread often assume that transmission mechanisms are fixed over time. However, social contagions--the spread of ideas, beliefs, innovations--can lose or gain in momentum as they spread: ideas can get reinforced, beliefs strengthened, products refined. We study the impacts of such self-reinforcement mechanisms in cascade dynamics. We use different mathematical modeling techniques to capture the recursive, yet changing nature of the process. We find a critical regime with a range of power-law cascade size distributions with non-universal scaling exponents. This regime clashes with classic models, where criticality requires fine tuning at a precise critical point. Self-reinforced cascades produce critical-like behavior over a wide range of parameters, which may help explain the ubiquity of power-law distributions in empirical social data.
\end{abstract}
\maketitle

\paragraph*{\textbf{Introduction}}
Cascades of beliefs, ideas, or news often show signs of criticality despite coming from various sources and spreading through different mechanisms \cite{notarmuzi2022universality}. The signature of criticality is a power-law tail in the cascade size distribution, scaling as $s^{-\tau}$. Cascade models predict this behavior at a precise critical point, the phase transition between a regime where all cascades eventually go extinct and another where they can grow infinitely. At this point, cascade models that follow a branching process structure universally predict a scaling exponent of $\tau=3/2$ \cite{harris1963theory}.  We call this critical exponent universal because, for a large family of spreading mechanisms, its value does not depend on the details of the model \cite{radicchi2020classes}. However, social media data show that cascade sizes can follow power-law distributions with scaling exponents much different from the prediction $\tau=3/2$. The size of message cascades and reply trees appear to decay much faster with scaling exponents of $\tau=3.9$ \cite{karsai2012universal} and $4$ \cite{nishi2016reply}, as do reposting cascades or recruitment to social movements both with exponents $\tau=2.3$ \cite{wegrzycki2017cascade, borge2011structural}, and many other data sources on platforms with exponents around $\tau=2$ \cite{notarmuzi2022universality}. The difference between the universality observed in models and the diversity of empirical results is yet unexplained.

Although cascade models vary, the vast majority of them use fixed mechanisms such that the same rules apply at every step of the cascade. For example, a new case of a disease produces infections through the same mechanism as previous cases. 
%Even when individual heterogeneity is included, the rules of the transmission model do not change as an outbreak progresses.
However, cascades of beliefs and ideas might be different. Beliefs can be self-reinforced \cite{zimmaro2025meta} or modulated by social interactions \cite{galesic2021integrating}. 
Ideas or products can be refined as they are transmitted from one person to another.

\begin{figure}
    \centering
    \includegraphics[width=0.78\columnwidth]{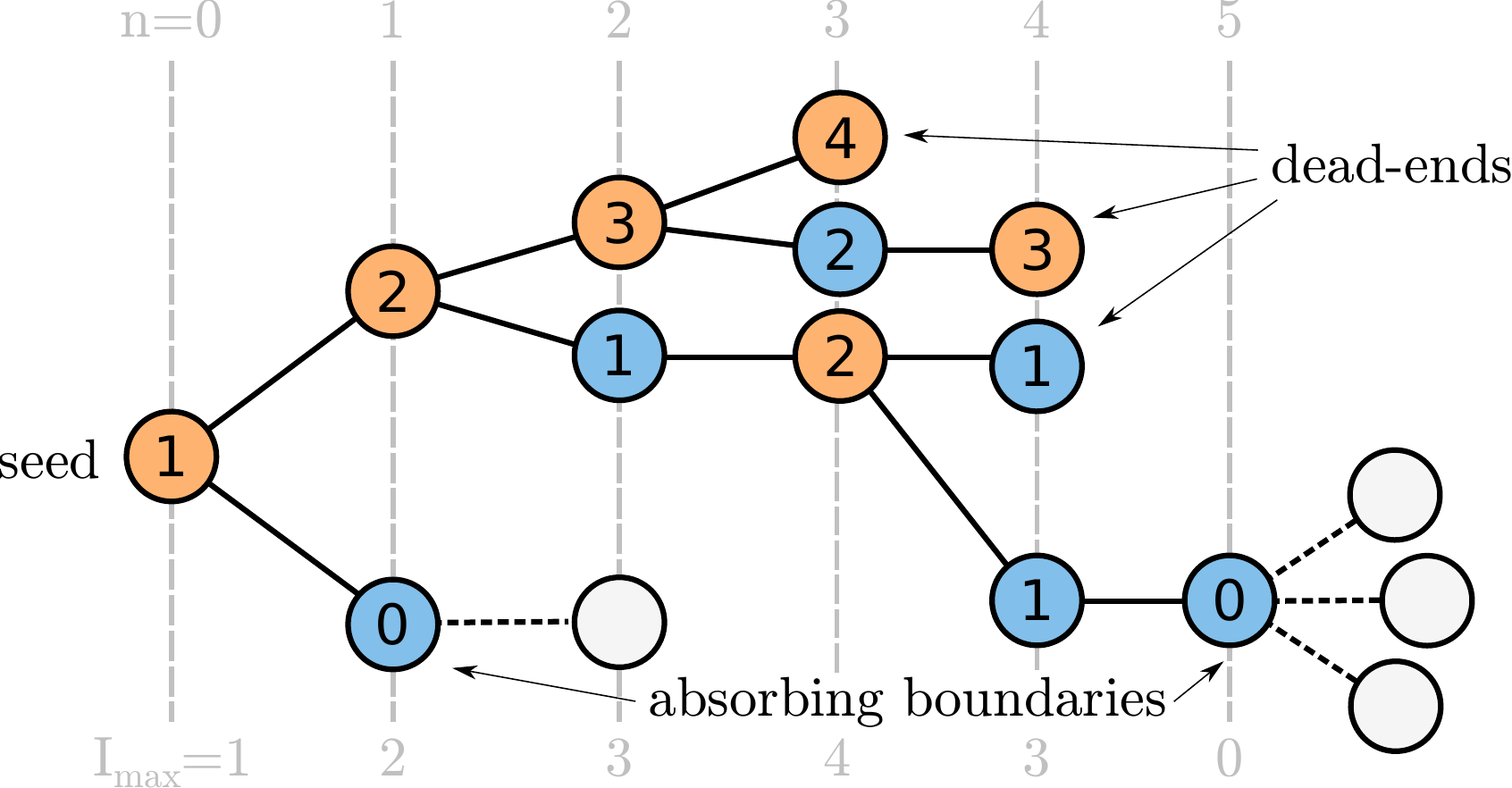}
	\caption{Schematic of a self-reinforcing cascade. Start with a seed of intensity one. At each generation $n$, the process gains a unit of intensity when reaching active neighbors (orange); or, loses a unit of intensity when reaching inactive neighbors (blue). Paths of the cascade end when they reach a node with no children (dead-end) or when the intensity falls to zero (absorbing boundary). The final cascade consists of all nodes where the process had positive intensity.}
    \label{fig:drawing}
\end{figure}

\paragraph*{\textbf{Self-Reinforcing Cascade (SRC) model}}
Imagine a cascading product like a meme, conspiracy theory, rumor, or a piece of software spreading in a population of agents. At every transmission step in the cascade, the product has the chance to independently improve with probability $p$ or get worse with probability $1-p$. This process can stop for two reasons: either the quality of the product drops to zero, or the agents sharing it cannot find others to pass it on to (see Fig.~\ref{fig:drawing}).

%To further explore the concepts of disease transmission models and the spread of ideas, beliefs, and behaviors, consider the fundamental difference in how these cascades evolve. Although disease transmission follows a mechanistic set of rules, the transmission of ideas and behaviors allows greater flexibility and adaptation. The reinforcement of beliefs or refinement of ideas introduces dynamic changes at each step, making these processes more unpredictable and context-dependent. 

As an example, consider open source projects where a seed piece of software is made available for others to fork and modify \cite{nyman2015understanding}. These modifications can either enhance or degrade the software, as well as its governance \cite{viseur2012forks, gamalielsson2014sustainability}. For instance, better code or governance might make the software more accessible and easier to adopt and update, while poorly written code or bad governance practices can make the software difficult to maintain, eventually leading to its abandonment \cite{coelho2017modern}. As a result, the quality of the software varies with each iteration, demonstrating the dynamic nature of such cascades.

%In recent years, in an effort to take into account the popularity-driven nature of many cascades, models have taken inspiration from Hawkes processes \cite{hawkes1971spectra}. Hawkes processes are autocorrelated point processes whose rate accelerates based on the number of recent event. Cascade models inspired by this mechanism can help explain critical cascades in two different universality classes: $\tau =3/2$ or $\tau =2$ \cite{notarmuzi2021percolation}.

%These feedback mechanisms are common and can lead to unpredictable and nonuniversal behavior. 
As a final and very different example, we note that we originally conceived the self-reinforcing mechanism as a model of forest fires gaining intensity as they burn trees but losing intensity as they traverse gaps in forest cover \cite{rednerWIP}. We explore this idea that cascades can deviate from universal classes when they can be amplified or attenuated as they spread.

%Although universality classes are widely accepted in many models of spreading phenomena, in practice, when examining real-world events such as the spread of misinformation or other context-dependent processes, the exponents we observe often vary significantly and do not necessarily fit neatly into these universal categories \cite{notarmuzi2022universality}. This suggests that the assumptions underlying these models may be oversimplified, particularly for more complex or evolving systems. Here, we explore the possibility that this deviation arises from a failure to account for self-reinforcing cascades, where the spread of information is amplified and sustained by feedback loops, leading to unpredictable and nonuniversal behavior.

SRC shares conceptual similarities with other models accounting for popularity-driven nature of many cascades, using self-exciting point processes \cite{notarmuzi2021percolation} or competition between cascades \cite{gleeson2014competition}. Cascade models based on these mechanisms can help explain two different universality classes, $\tau =3/2$ and $\tau =2$ \cite{notarmuzi2021percolation}, and interpolate between \cite{gleeson2014competition}.

More generally, SRC are a cascade perspective on killed branching random walks where certain results are known for its critical point in a continuous limit \cite{kesten1978branching} and bounds on its critical behavior in the discrete case \cite{aidekon2013precise}. Here, we provide exact recursive solutions, static closed-form expressions for the expected cascade sizes and their critical point, and a dynamic analysis of cascade intensity. Beyond theoretical contributions, our results illustrate how diverse scaling exponents can easily be observed in cascades and we provide estimates for the exponents produced by SRC.

%Our model takes a simple approach to addressing this problem. Imagine something (e.g., a meme, conspiracy theory, rumor, code) cascading in a population of agents and that at each iteration of the cascade it either improves with probability $p$ or worsens with probability $1-p$. Any chain of transmission within that cascade can end when the quality or intensity of the cascade falls to zero \textit{or} when agents fail to find neighbors to continue the process (see Fig.~\ref{fig:drawing}).

\paragraph*{\textbf{Recursive solution}}

%We call self-reinforcing cascades (SRC) a directed percolation process that gains a unit of intensity when encountering occupied sites and that can thereafter spend one of those units to activate unoccupied sites. Occupied sites are interpreted as receptive to the cascade (receptive to a belief, idea, joke, or ability to improve a product), and unoccupied sites are non-receptive (in need of convincing or agents that worsen the product).

%We interpret the model as a dynamical process starting from a random seed and extending to neighbors. The cluster starts with intensity one, and its intensity grows by one when the cluster grows to an occupied neigbor. Positive intensity allows the cluster to reach unoccupied nodes but thereby decreasing its intensity by one. Any branch of the cluster ends when they reach an unoccupied node with zero intensity (i.e., we do not allow negative intensity nodes). 

%\subsection{Probability generating functions}

Mathematically, we consider a general branching structure for contacts within the population. At any node, we define $G(x) = \sum_b \pi_b x^b$ as the probability generating function (PGF) for the number of ``children'' neighbors (occupied or not) of that node, $\pi_b$ being the probability of branching into exactly $b$ children \cite{wilf1994generating}. The probability $\pi_0$ is equal to the probability that any node in the cascade is a dead-end without children, one of two ways for a chain of transmission to end. Importantly, and in line with recent empirical findings \cite{gleeson2020branching}, the branching number $b$ is drawn independently and identically at each node. However, intensity-dependent distributions for the branching number is a straightforward extension of the model (explored in Supplemental Material). %Importantly, we generally require $p_0 \neq 0$  such that nodes can be dead-ends and the process is not guaranteed to continue forever.
%In $G(x)$, $x$ is a counting variable whose power counts the number of potential children.

We pick the first node on this structure to start a cascade of intensity 1 (more generally, $I_0$). Any potential children will be either receptive to the process with probability $p$, and continue the process with intensity 2; or non-receptive, such that they end their branch of the cascade by reaching intensity 0. In the next step, children with non-zero intensity in the last step (if any) can recruit their own receptive children (if any) to continue the process with intensity 3; or convince their non-receptive children (if any) to continue the process with intensity 1. In general, intensity increases when the cascade spreads to receptive nodes and decreases when it spreads to non-receptive ones. Any branch of the cascade dies either when it reaches a dead-end or when its intensity goes to zero.

We can solve this process using a self-consistent recursive solution. Let $H_1(x)$ be the PGF for the cascade size distribution of a node of intensity 1 \cite{newman2001random}. Since the root node is part of the cascade, $H_1(x)$ has to be proportional to $x$ to count that node. After that, every possible children generated by $G(x)$ is either receptive with probability $p$ or non-receptive otherwise. Receptive children will have intensity 2 and produce cascades whose size is generated by $H_2(x)$, while non-receptive children will have intensity 0 and thus lead to trivial cascades of size 0, i.e, $H_0(x) = x^0 = 1$. We can therefore write $H_1(x) = xG\left[pH_2(x)+(1-p)\right]$
to define a recursive self-consistent equation. In this equation, we use the fact that a cascade produced by a node is the sum of the cascades produced by its children, and that the PGF for the sum of a variable number of independent random variables is the composition of the PGFs.

More generally, we can write
\begin{equation}
H_k(x) = xG\left[pH_{k+1}(x)+(1-p)H_{k-1}(x)\right]
\label{eq:recursion}
\end{equation}
as non-receptive nodes decrease the intensity but do not necessarily end the process if $k>1$. To close the second-order recursion in Eq.~(\ref{eq:recursion}) we need another boundary condition besides $H_0(x)=1$. To this end, we define $\bar{k}$ as the maximum intensity allowed, implemented by requiring $H_{\bar{k}}(x) = H_{\bar{k}+1}(x)$.

All closed-form solutions presented below hold in the limit $\bar{k}\to\infty$, i.e., for cascades with no upper bound on intensity. All numerical simulations are performed in this limit. In practice, integrating Eq.~(\ref{eq:recursion}) requires choosing a finite $\bar{k}$. However, as shown in the following, cascades seeded at an intensity enough smaller than $\bar{k}$ are insensitive to the latter, which is thus effectively indistinguishable from $\bar{k}\to\infty$.

To solve Eq.~(\ref{eq:recursion}), we iterate it for multiple values of $x$, up to $\bar{k}$ (we use 100), until all values converge to within a certain precision threshold (we use $10^{-12}$).
Once a fixed point has been reached, we can extract $H_1(x)$, which generates the cascade size distribution from the seed (assuming $I_0 = 1$). We can also obtain the probability that a cascade is supercritical and never ends as $1-H_1(1)$, being infinite cascades not accounted for in Eq.~(\ref{eq:recursion}). Methods are detailed in our Supplemental Material (SM) \cite{supp} and our calculations are available online \cite{data}.

%Looking at the recursive structure of Eq.~(\ref{eq:recursion}), we can highlight how the SRC process is a mixture of a branching process with a biased random walk of intensity. If we fix the number of potential offsprings to one such that $G(x) = x$, then our recursion solves for the number of steps until a biased random walk starting at $I_0$ reaches the absorbing boundary, as it makes either a step up with probability $p$ or a step down with probability $1-p$. For more general $G(x)$, SRCs are therefore similar to a biased random walk that dies after every step but generates children through $G(x)$; and we are again calculating the distribution of steps until every member of the population has reached the absorbing state of zero intensity. Similarly, the process can fall back on classic directed percolation if we set $H_{k>1}(x) \equiv H_1(x)$.

\begin{figure}
	\centering
	\includegraphics[width=0.95\columnwidth]{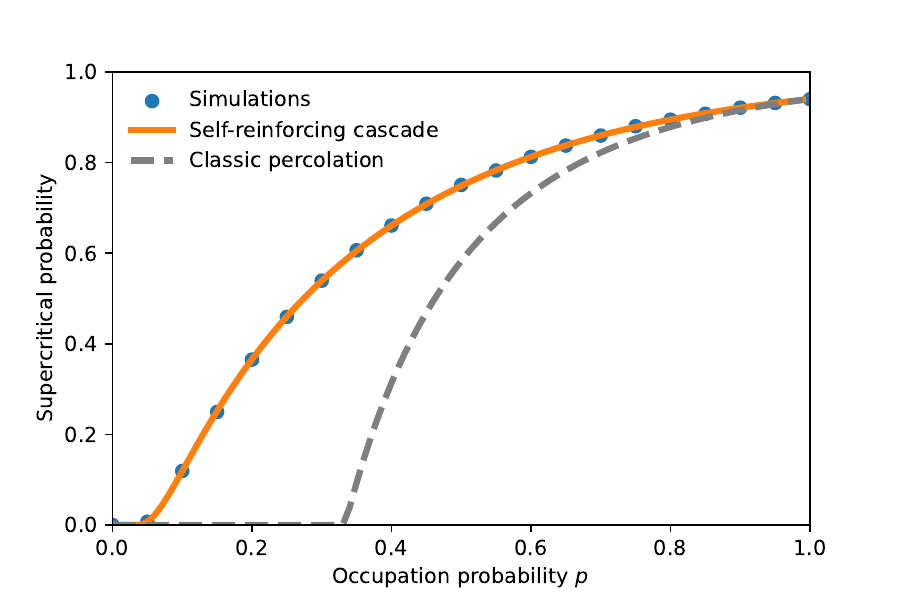}
	\caption{Phase transitions of SRC and directed percolation on Poisson trees of average branching number $\ell=3$. For the SRC, we compare our recursive exact solution based on Eq.~(\ref{eq:recursion}) to simulations. The critical point marking the emergence of a supercritical cascade is at $p=1/\ell=1/3$ for percolation and at $p = (1-2\sqrt{2}/3)/2 \approx 0.0286$, as computed from Eq.~(\ref{eq:pc}), for the SRC.}
    \label{fig:transition}
\end{figure}

\paragraph*{\textbf{Critical point}}

\begin{figure}
	\centering
	\includegraphics[width=0.95\columnwidth]{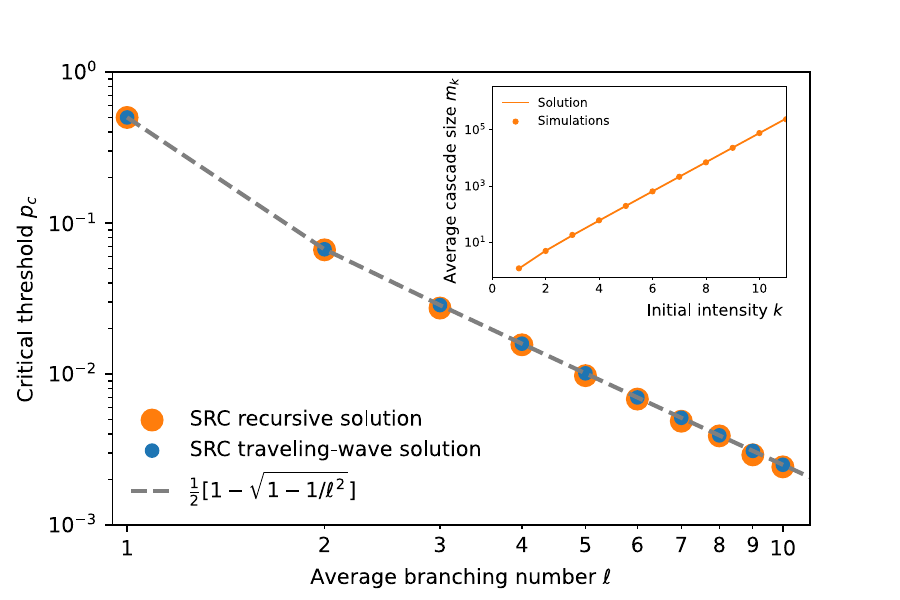}
	\caption{Critical threshold $p_c$ of SRC on Poisson trees of different average branching number $\ell$. Results are obtained by solving the exact recursion in Eq.~(\ref{eq:recursion}), the explicit solution in Eq.~(\ref{eq:pc}), and the critical condition of the traveling wave in Eq.~(\ref{eq:tv_imax}). The results match up to the numerical precision at which we solve the recursion. The inset validates the explicit solution in Eq.~(\ref{eq:m_k}) for the expected cascade size $m_k$, comparing it with $10^4$ simulations per value of initial intensity, performed at $p=0.01$ and $\ell = 3$.}
    \label{fig:thresholds}
\end{figure}

\begin{figure*}
	\centering
	\includegraphics[width=0.95\columnwidth]{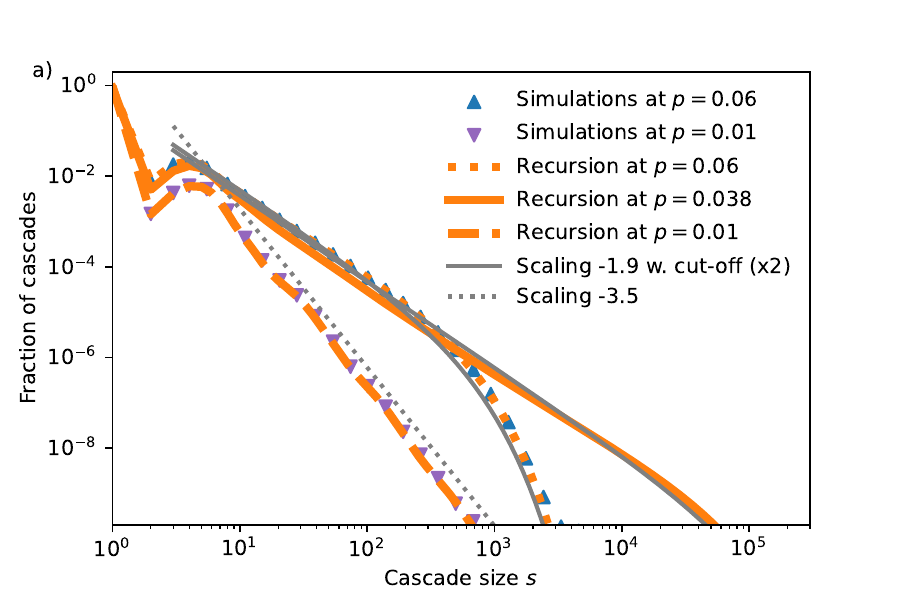}
	\includegraphics[width=0.95\columnwidth]{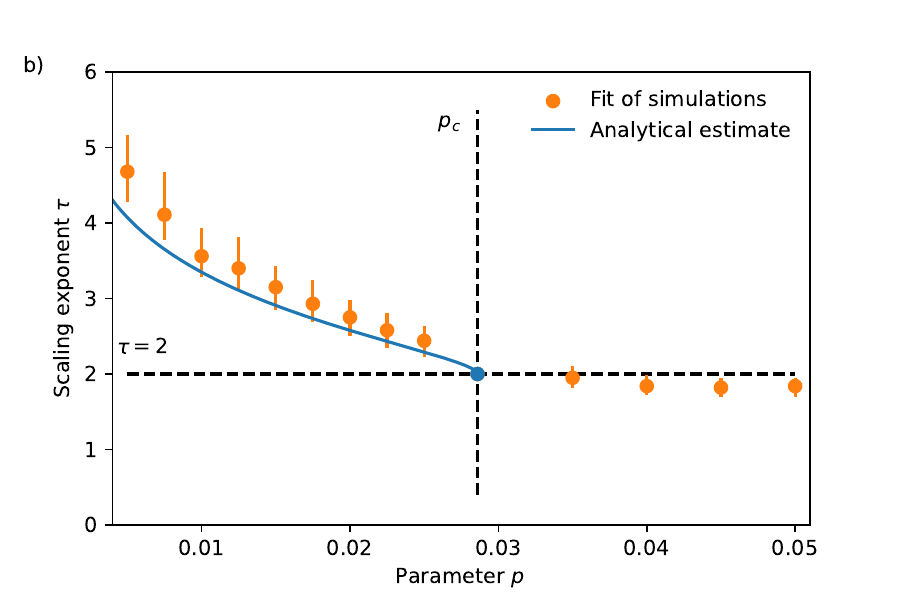}
	\caption{Extended critical behavior around the critical point $p_c$ for a Poisson tree of $\ell = 3$ ($p_c\approx 0.0286$). (a) Cascade size distributions for $p$ above and below $p_c$. Above $p_c$, we find a scaling relationship with exponential cutoff $s^{-\tau(p)} \times e^{-s/\bar{s}(n_c(p))}$ based on the critical generation $n_c(p)$ given in Eq.~(\ref{eq:n_c}) if $p$ is close to $p_c$. %Specifically, we show results for $p=0.038$, for which $n_c \approx 10.66$ and $\tau \approx 1.78$. 
    Below $p_c$, we find arbitrarily steep power-law decays as a function of $p$; for instance, $\tau\approx 3.5$ for $p=0.01$. Results from $10^8$ simulations are reported for two $p$ values with logarithmic binning. The recursion is exact. (b) Scaling exponents versus $p$ obtained with our approximate solution, Eq.~(\ref{eq:taup}), and by fitting power-law tails to our exact solution from recursion. Goodness-of-fit is evaluated with the Kolmogorov–Smirnov (KS) statistic \cite{clauset2009power}; markers show its minima and lines show an acceptable range (KS $< 0.05$). }
    \label{fig:clusters}
\end{figure*}

We first look at the phase transition of SRC in Fig.~\ref{fig:transition}. 
We assume that the number of children nodes is drawn from a Poisson distribution $\pi_b = \ell^b e^{-\ell}/b!$ of mean $\ell \geq 1$. 
The self-intensifying mechanism greatly reduces the critical point of the process. For an average branching number $\ell = 3$, we find $p_c$ at $(1-2\sqrt{2}/3)/2 \approx 0.0286$ instead of $1/\ell = 1/3$ for the emergence of a giant connected component (infinite-size) cascade in a random network \cite{molloy1995critical}.

We now derive a closed-form solution for the critical point. 
%Unlike standard percolation, which is solved by a single self-consistent equation, Eq.~(\ref{eq:recursion}) yields an infinite system of coupled equations. 
To gain some insights into the expected behavior of Eq.~(\ref{eq:recursion}), we rewrite the system as a recursion over the expected cascade size $m_k(p)$ when starting at a node of intensity $k$ for a given $p$. To calculate $m_k(p)$, we average the cascade size distribution by taking the derivative of $H_k(x)$ and evaluating at $x=1$ (with $H_k(1) = 1$ in the subcritical regime), to obtain
\begin{equation}
    m_k(p) = 1+\ell p m_{k+1}(p) + \ell (1-p)m_{k-1}(p) \; .
\end{equation}
%The homogeneous part of this recursion, ignoring the constant, has solutions of the form $m^{(0)}_k = cA^k$. Solving for $A$, we then extend the solution to $m_k = cA^k +b$ and solve for all remaining parameters knowing that 
This is a second-order linear difference equation with boundary conditions $m_1(p) = 1 + \ell p m_2(p)$ and $m_{\bar{k}}(p) = 1+\ell p m_{\bar{k}}(p) + \ell (1-p)m_{\bar{k}-1}(p)$. In the limit of large $\bar{k}$, we obtain the exact solution (see SM for derivation \cite{supp})
\begin{equation}
    m_k(p) = \frac{1}{\ell-1}\left\{\left[\frac{1-\sqrt{1-4p(1-p)\ell^2}}{2p\ell}\right]^k - 1\right\} \; .
    \label{eq:m_k}
\end{equation}
%The critical point of the process can then be found as the highest value of $p$ such that $m_k(p) \in \mathbb{R}$. 
%More formally,
We calculate the critical point $p_c$ of the process as the value of $p$ where the susceptibility of the system diverges, such that $dm_k/dp \rightarrow \infty$. We find
\begin{equation}
    p_c= \frac{1}{2}\left(1-\sqrt{1-\ell^{-2}}\right) 
    \label{eq:pc} \; .
\end{equation}
Figure \ref{fig:thresholds} validates the exact closed-form expressions above by comparing Eq.~(\ref{eq:pc}) against our other solutions for $p_c$ and Eq.~(\ref{eq:m_k}) against numerical simulations (inset plot).

\paragraph*{\textbf{Temporal behavior}}
%\section{Extended critical behavior and a traveling wave solution}

To get a better idea of the behavior of the system around the critical point, we 
%rely on known results for extremal paths on trees \cite{majumdar2000extremal} to 
analyze the depth of cascades and the associated temporal dynamics of intensity. 
We write a recursion for $Q_k(n)$, the cumulative probability that a cascade generated by a node of intensity $k$ has depth not larger than $n$,
\begin{equation}
    %D_k(n) = 1\! -\! G\left[1\!-\!pD_{k+1}(n\!-\!1)\!-\!(1\!-\!p)D_{k-1}(n\!-\!1)\right]
    Q_k(n) = G\left[pQ_{k+1}(n-1)+(1-p)Q_{k-1}(n-1)\right] \; ,
    \label{eq:depth_recursion}
\end{equation}
with initial condition $Q_0(n) = 1-\delta_{n,0}$.
Crucially, solving for the tail $D_k(n)=1-Q_k(n)$ reveals that, independently of the initial intensity, the tail of the depth distribution decays exponentially for large $n$ as $D(n) \sim \exp(-n \gamma_\ell(p))$, with $\gamma_\ell(p) = -\ln\sqrt{4p(1-p)\ell^2}$. See SM for details \cite{supp}.

At each generation $n$, we can calculate the expected maximal number of positive steps in intensity $P_{\textrm{max}}(n,p)$ (i.e., the number of receptive nodes met) over all paths. 
%Previous work did not consider absorbing boundaries as we do at intensity equal to zero \cite{majumdar2000extremal}, but we assume that if there exists a path without the absorbing boundary where the number of positive steps goes to infinity, then there exists a path that can avoid extinction despite the absorbing boundary. 
To solve for the dynamics of $P_{\textrm{max}}(n,p)$ we define the cumulative probability $R_n(x) = \textrm{Prob}(P_{\textrm{max}}(n,p) \leq x)$, with initial condition $R_0(x) = \mathbf{1}_{x\geq 0}$. This obeys the recursion
\begin{equation}
    R_{n+1}(x) = G\left[pR_n(x-1) + (1-p)R_n(x)\right] \; .
    \label{eq:tw_rec}
\end{equation}
We then use a traveling wave Ansatz $R_n(x) = \tilde{R}(y=x-v_{\textrm{max}}n)$. We linearize Eq.~(\ref{eq:tw_rec}) in the region far ahead of the front and look for an exponential solution, $1-\tilde{R}(y) \sim e^{-\mu y} \ll 1$. We find that the values of the velocity $v_{\textrm{max}}$ and the decay exponent $\mu$ of the centered cumulative probability function $\tilde{R}(y)$ are related and found by solving a transcendental equation,
\begin{equation}
v_{\textrm{max}}(p) = \frac{1}{\mu}\ln\left[\ell(1-p)+\ell p e^{\mu}\right]  = \dfrac{p e^{\mu}}{1-p+p e^{\mu}}\; .
\label{eq:der=0}
\end{equation}
\begin{comment}
\begin{equation}
v_{\mu}(p) = \frac{1}{\mu}\ln\left[\ell(1-p)+\ell p e^{\mu}\right] \; ,
\end{equation}
as parametrized by $\mu$. The value of the latter is fixed by the principle of velocity selection so to minimize $v_{\textrm{max}}(p)$~\cite{majumdar2000extremal}, as given by the solution to the following transcendental equation:
\begin{equation}
\ln\left[\ell(1-p)+\ell p e^{\mu}\right] - \dfrac{p\mu e^{\mu}}{1-p+p e^{\mu}} = 0 \; .
\label{eq:der=0}
\end{equation}
The first term is just $\mu v_{\mu}(n,p)$ and we write the selected front velocity (and related critical point) as
\end{comment}
%We solve the equation for $\mu$ and thus obtain the front velocity and related critical point as
%\begin{equation}
%v_{\textrm{max}}(p) = \dfrac{pe^{\mu}}{1-p+pe^{\mu}}\ \ \Rightarrow\ \ \dfrac{p_ce^{\mu}}{1-p_c+p_ce^{\mu}}  = \frac{1}{2} \; .
%\label{eq:critical}
%\end{equation}
The first equality comes from the linearization of Eq.~(\ref{eq:tw_rec}); the second one from solving for the minimum value of the corresponding speed, in accordance to the principle of velocity selection \cite{majumdar2000extremal}.

The traveling wave Ansatz comes with a universal logarithmic correction \cite{bramson1978maximal,brunet1997shift,brunet2016some}, such that we have $P_{\textrm{max}}(n,p) = v_{\textrm{max}}(p)n + \frac{3}{2\mu}\log n$. %, where the velocity $v_{\textrm{max}}(p)$ can be interpreted as the probability of making a positive step along a branch of maximal intensity. 
Assuming an initial intensity $I_0$, the expected maximal intensity $I_{\textrm{max}}(n,p)$ after $n$ generations will be $I_0$ plus the difference between positive and negative steps after $n$ generation; that is,
\begin{align}
I_{\textrm{max}}(n,p) & = P_{\textrm{max}}(n,p) - (n-P_{\textrm{max}}(n,p)) + I_0 \nonumber \\
 & = 2P_{\textrm{max}}(n,p) - n  + I_0\; .
\end{align}
%since the process is expected to have reached $P_{\textrm{max}}(n,p)$ receptive nodes that increased intensity by 1 and therefore $n-P_{\textrm{max}}(n,p)$ non-receptive nodes that decreased intensity by 1.
By definition of $p_c$, $I_{\textrm{max}}(n,p)$ should diverge as $n\rightarrow \infty$ for $p > p_c$ and go to zero for $p < p_c$. Therefore, at $p_c$,
\begin{equation}
    \lim_{n\rightarrow \infty}\frac{dI_{\textrm{max}}}{dn} = 2v_{\textrm{max}}(p_c) - 1 = 0\ \ \Rightarrow\ \ v_{\textrm{max}}(p_c) = 1/2 \; .
    \label{eq:tv_imax}
\end{equation}

We again find $p_c$ by imposing $v_{\textrm{max}}(p_c) = 1/2$ in Eq.~(\ref{eq:der=0}). %and substituting it in the logarithm of Eq. (\ref{eq:der=0}). 
%This derivation yields the same solution as Eq.~(\ref{eq:pc}), as shown in Fig.~\ref{fig:thresholds}.% validates this solution against the recursive solution obtained from Eq.~(\ref{eq:recursion}) and the explicit one from 
%(See Appendix B for a fourth derivation of the critical point.)

% \begin{figure}[]
% 	\centering
% 	\includegraphics[width=0.9\columnwidth]{figure5_combined.pdf}
% 	\caption{Distributions of cluster size $s$ above and below the critical point, validated with exact simulations. Above the critical point, we find a scaling relationship with exponential cutoff $s^{-2} \times \textrm{exp}(-s/\ell^{n_c})$ based on the generation $n_c$ at which the traveling wave solution for $p=0.06$ crosses the critical condition. Below the critical point, we can find arbitrarily steep power-law decay.}
%  \label{fig:super} 
% \end{figure}

%We are now interested in the cascade size distribution around the critical point, 
\paragraph*{\textbf{Extended critical behavior}}
Figure \ref{fig:clusters} shows that below $p_c$, the cascade size distributions of the SRC feature steep power-law tails with tunable scaling exponents $\tau>2$. The exponent decreases when $p$ increases until reaching $\tau = 2$ at $p = p_c$, as required for the expected cascade size to diverge. %For larger values of $p$, we find that a robust power-law behavior with exponential cutoff exists well above $p_c$. %This is very different from what is observed in classic directed percolation, where power-law cascade distributions are found only at precisely the percolation threshold. 
This behavior becomes even more persistent for more heterogeneous branching distributions \cite{supp}.

Why do we find a critical-like scaling off the critical point? By combining all of our analyses, we can: i. estimate the scaling exponents in the subcritical regime, ii. derive the critical exponent, and iii. characterize the exponential cutoffs in the supercritical regime. Importantly, the presence of power-law tails in the subcritical regime has recently been proven in the context of killed branching random walks in Ref. \cite{aidekon2013precise} without explicit solutions for the scaling exponents.
%Mechanistically, the power-law distributions occur because we are looking at cascade size, which is exponentially related to cascade intensity as per Eq.~(\ref{eq:m_k}), and the distribution of intensity was found to have an exponential tail as per our traveling wave solution. This combination of exponentials is a known mechanism to produce power-law tails \cite{newman2005power}.

Mathematically, cascade size increases exponentially with the intensity $k$ as $s\sim \textrm{exp}\left(ak\right)$, where $a$ can be readily identified from Eq.~(\ref{eq:m_k}). On the other hand, according to our traveling wave solution, the distribution of the maximal intensity $I_\textrm{max}(n)$ at a given generation has an exponential tail. Based on the tail of cascade depth $D_k(n)=1-Q_k(n)$ from Eq.~(\ref{eq:depth_recursion}), we can integrate over exponentially decaying cascade depths. The tail of cascade sizes is then generated by the exponential of a quantity, $I_\textrm{max}$, distributed with an exponential tail $e^{-bI_\textrm{max}}$. This combination of exponentials is a known mechanism to produce power-law tails \cite{newman2005power}. As detailed in SM \cite{supp}, we can thus combine the rates $a$ and $b$ to estimate the scaling exponent $\tau(p) = 1+b/a$ as
\begin{equation}
    \tau(p) \approx 1+\frac{\mu \left[1-v_\text{max}(p)\right] +\ln\sqrt{4p(1-p)\ell^2}}{\ln\left(1-\sqrt{1-4p(1-p)\ell^2}\right)-\ln(2p\ell)}
\label{eq:taup}
\end{equation}
when $p<p_c$, as shown in Fig.~\ref{fig:clusters}(b). Notice that, consistent with a derivation shown in our SM using only the traveling wave at $p=p_c$, Eq.~(\ref{eq:taup}) correctly predicts $\tau(p_c)=2$.

%As the parameter $p$ approaches $p_c$, the expected cascade size $m_1(p\rightarrow p_c)$ diverges, and so does our solution for the exponential distribution of maximal intensity per cascade. The distribution of maximal intensity per cascade therefore undergoes a classic phase transition, its distribution going from exponential at $p<p_c$ to a power law at $p_c$. The extended regime of power-law size distributions is a consequence of the exponential relationship between maximal intensity and expected size.

Moving to the supercritical regime, we can characterize the distributions further using the universal logarithmic correction to $P_{\textrm{max}}(n,p)$. This correction implies that the process is expected to spend $n_c(p)$ generations close to extinction before the linear growth of the wave front dominates (validated in Fig. 5). At small $n$, the critical condition reads $dI_{\textrm{max}}/dn = 2v_{\textrm{max}}(p) - 1 - 3/(\mu n) = 0$, and we find
\begin{equation}
n_c(p) = \frac{3}{\mu (2v_{\textrm{max}}(p)-1)} \; .
\label{eq:n_c}
\end{equation}

\begin{figure}[t!]
	\centering
	\includegraphics[width=0.95\columnwidth]{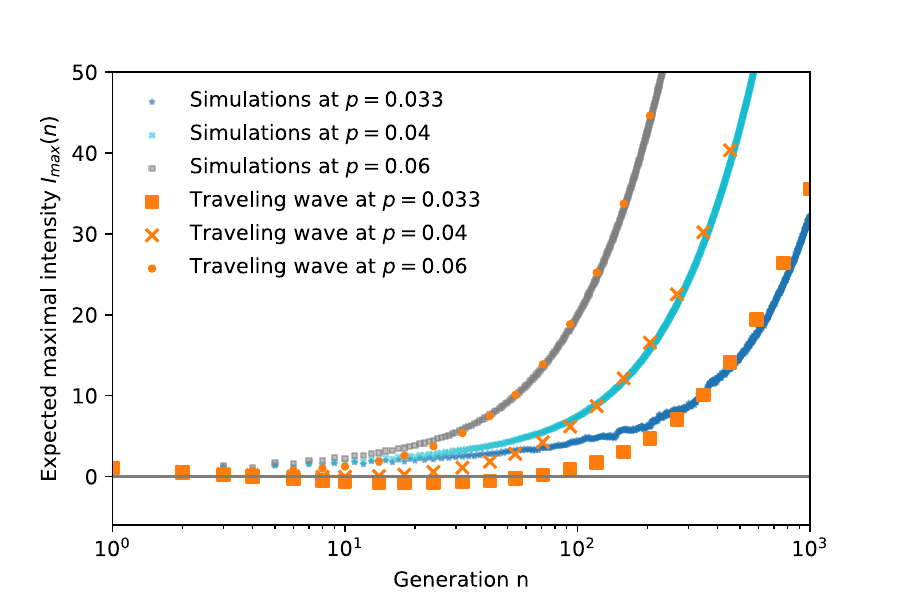}
	\caption{Expected maximal intensity over generations $n$ produced by the logarithmically-corrected $I_{\textrm{max}}(n,p)$ for different values $p > p_c$. We compare the solution with the average maximal intensity observed in at least $10^6$ given the process is not extinct at generation $n-1$. By definition, surviving cascades in simulations are always at intensity greater than zero. Nonetheless, the traveling wave solution captures the general long-time behavior of $I_{\textrm{max}}(n,p)$.}
    \label{fig:crit_traj}
\end{figure}

%Fixed $p>p_c$, such that $v_{\textrm{max}}(p) > 1/2$, $dI_{\textrm{max}}/dn < 0$ for all the first generations $n < n_c(p)$, implying that the cascades behave subcritically until then. 
%We expect a power-law behavior for cascades of size not larger than $\bar{s}(n_c) = (\ell^{n_c} - 1)/(\ell - 1)$, the expected cascade size reachable by generation $n_c$. 
%If $n$ reaches $n_c$, then $dI_{\textrm{max}}/dn > 0$, and the typical exponential behavior above the percolation threshold should be recovered. 
As is the case with most cascade models, we expect an exponential cutoff on the cascade size distribution which is here imposed by the critical generation $n_c$ and should thus be of order $\bar{s}(n_c) = (\ell^{n_c} - 1)/(\ell - 1)$, the maximum size a cascade can reach by generation $n_c$. We estimate the cascade size $s$ to be distributed as a power-law with exponential cutoff $s^{-\tau(p)} \times e^{-s/\bar{s}(n_c(p))}$. As Fig.~\ref{fig:clusters}(a) shows for $p = 0.038$, for which $n_c \approx 10.66$ and $\tau \approx 1.9$, our estimation is in excellent agreement with the exact results from recursion. Nonetheless, as the value of $p$ increases, $n_c$ decreases, and additional correction terms would eventually come into play to push the cutoff to higher values (first with a non-universal term of order $1/\sqrt{n}$) \cite{brunet2016some}. We illustrate this using $p=0.06$ in Fig.~\ref{fig:clusters}(a), where Eq.~(\ref{eq:n_c}) predicts $n_c \approx 3.91$, yet we obtain a much better fit with the value $n_c = 6$ used in the figure. %The values of $\tau(p>p_c)$, which appear to decrease after $p_c$, also remain an open problem.%Our estimated cutoff thus offers a lower bound, such that the scaling behavior is observed for supercritical values of $p$ higher than expected from $\bar{s}(n_c(p))$. 
However, the exact value of $\tau(p>p_c)$, and whether this value truly deviates from $\tau(p_c) = 2$, remains an open problem.

%For cascades that die at a finite depth $n$ where the logarithmic correction term can not be ignored, larger $p>p_c$ values might not satisfy the conditions for supercriticality for all early generations $n$. These parameters that cross the critical condition at some point as $n$ increases might, therefore, still show the signature of criticality. At a given $p\neq p_c$ value, we find that the critical condition is satisfied at some effective generation $n_c(p)$ before the condition becomes supercritical (greater than zero) for $n>n_c(p)$. This, in turn, might impose an exponential cutoff on the cluster size distribution.  

%In Fig.~\ref{fig:crit_traj}, we look at the expected maximal intensity $I_{\textrm{max}}(n,p)$ versus $n$ for some $p>p_c$ to further validate the traveling wave solution. %illustrate the rationale behind Eq.~(\ref{eq:n_c}). %In a nutshell, the long-time behavior of the expected maximal intensity is what determines the critical point, but its transient behavior characterizes the bulk of that distribution.

\paragraph*{\textbf{Discussion}}
The self-reinforcing cascade process is a parsimonious model to capture the fact that the strength of individual beliefs or the quality of products may vary and thus influence their ability to spread further. This variability is aligned with real-world phenomena, where not all individuals or contents are equally influential in the transmission of ideas or behaviors. %Future work could investigate the impact of having the branching function $G(x)$ depend on intensity as $G(x,k)$, but our minimal model highlights the drastic difference between standard branching process and SRC even without this coupling. 
With this simple mechanism, the SRC model can produce a wide range of scaling behaviors for cascade size distributions, whereas classic cascade models are constrained by a unique and universal scaling exponent obtained only at a precise critical point. 

The shape and statistics of SRC are surprising in two ways. One, they follow fat-tailed cascade size distribution in their subcritical regime. Two, the tail of their depth distribution remains exponential even at the critical point. Thus, subcritical and critical SRC are short but very broad compared to classic cascade models. The unique shape of SRC could help identify signatures of varying cascade intensity or quality in empirical data. We could expect this mechanism to appear in cascades where content is able to change (e.g., mutations in epidemics or personalized social media posts) as opposed to cascade with fixed content (e.g., cascade of clones, or reposting in social media), and SRC could provide more realistic models in these cases. In addition, the search remains open for a general model that provides anomalous exponents for both size and depth distributions as found empirically \cite{notarmuzi2022universality}.

%While self-similar or isotropic spreading rules are mathematically convenient, providing straightforward solutions and modeling frameworks, self-reinforcing cascades can offer advantages that warrant further study. The flexibility and richness of their outcome suggest that they are better suited to capture the complexities of real-world social contagions. 

We outlined several important properties of self-reinforcing cascades and proposed a few analytical approaches to better understand these processes: exact probability generating functions for cascade size, exact recursions over cascade depth, and the traveling wave technique for cascade intensity. Combining all three approaches, we were able to characterize the different scaling behaviors produced by the model. Altogether, this effort may provide a useful framework for researchers and practitioners seeking to understand cascading behavior in complex real-world systems. 

\paragraph*{\textbf{Acknowledgments}}
The authors acknowledge financial support from The National Science Foundation awards \#2419733 (L.H.-D. and G.B.) and \#2242829 (J.L. and G.B.) as well as from Science Foundation Ireland under Grant number 12/RC/2289 P2 (J.P.G.). The research was supported by International Partnerships and Programs at the University of Vermont as a Global Catalyst Research Partnership Award.

\end{document}